\documentclass[a4paper,fleqn]{cas-sc}

\usepackage[authoryear]{natbib}
\usepackage{cancel}
\usepackage{lineno}

\def\tsc#1{\csdef{#1}{\textsc{\lowercase{#1}}\xspace}}
\tsc{WGM}
\tsc{QE}

\newcommand\T{{\rm T}}

\begin{document}
\let\WriteBookmarks\relax
\def\floatpagepagefraction{1}
\def\textpagefraction{.001}

\shorttitle{Seasonal Outgassing on Dark Comets}    

\shortauthors{Taylor et al.}  

\title [mode = title]{Seasonally Varying Outgassing as an Explanation for Dark Comet Accelerations}

\author[1,2]{Aster G. Taylor}[orcid=0000-0002-0140-4475]

\fnmark[1]
\fntext[1]{Fannie and John Hertz Foundation Fellow}

\cormark[1]
\cortext[1]{Corresponding author: Aster G. Taylor, \texttt{agtaylor@umich.edu}}

\author[3]{Davide Farnocchia}[orcid=0000-0003-0774-884X]
\author[4]{David Vokrouhlick\'y}[orcid=0000-0002-6034-5452]
\author[5]{Darryl Z. Seligman}[orcid=0000-0002-0726-6480]
\author[6,7]{Jordan K. Steckloff}[orcid=0000-0002-1717-2226]
\author[8]{Marco Micheli}[orcid=0000-0001-7895-8209]

\affiliation[1]{organization={Dept. of Astronomy and Astrophysics, University of Chicago},
            city={Chicago},
            postcode={60637}, 
            state={IL}}
\affiliation[2]{organization={Dept. of Astronomy, University of Michigan},
            city={Ann Arbor},
            postcode={48109}, 
            state={MI}}
\affiliation[3]{organization={Jet Propulsion Laboratory, California Institute of Technology},
            addressline={4800 Oak Grove Dr.}, 
            city={Pasadena},
            postcode={91109}, 
            state={CA},
            country={USA}}
\affiliation[4]{organization={Institute of Astronomy, Charles University},
            addressline={V Hole\v{s}ovi\v{c}k\'ach 2 }, 
            city={CZ-18000 Prague 8},
            country={Czech Republic}}
\affiliation[5]{organization={Dept. of Astronomy and Carl Sagan Institute, Cornell University},
            addressline={122 Sciences Dr.}, 
            city={Ithaca},
            state={NY},
            postcode={14853}, 
            country={USA}}
\affiliation[6]{organization={The Planetary Science Institute},
            city={Tuscon},
            state={AZ},
            country={USA}}
\affiliation[7]{organization={Department of Aerospace Engineering and Engineering Mechanics, The University of Texas at Austin},
            city={Austin},
            state={TX},
            country={USA}}
\affiliation[8]{organization={ESA NEO Coordination Centre},
            addressline={Largo Galileo Galilei 1}, 
            city={I-00044 Frascati (RM)},
            country={Italy}}

\begin{abstract}
Significant nonradial, nongravitational accelerations with magnitudes incompatible with radiation-driven effects have been reported in seven small, photometrically inactive near-Earth objects. Two of these objects exhibit large transverse accelerations (i.e., within the orbital plane but orthogonal to the radial direction), and six exhibit significant out-of-plane accelerations. Here, we find that anisotropic outgassing resulting from differential heating on a nucleus with nonzero spin-pole obliquity, averaged over an eccentric orbit, can explain these accelerations for most of the objects. This balanced outgassing model depends on three parameters --- the spin pole orientation (R.A. and Dec.) and an acceleration magnitude. For these ``dark comets" (excepting 2003 RM), we obtain parameter values that reproduce the observed nongravitational accelerations. We derive formulae for the component accelerations under certain assumptions for the acceleration scaling over heliocentric distance. Although we lack estimates of these objects' spin axes to confirm our values, this mechanism is nevertheless a plausible explanation for the observed accelerations, and produces accurate perturbations to the heliocentric motions of most of these objects. This model may also be applied to active objects outside of the dark comets group.
\end{abstract}




\maketitle

\section{Introduction} \label{sec:intro}

Classically, Solar System small bodies are divided into two populations --- asteroids and comets --- based on appearance and detectable mass-loss. Comets are composed of a mixture of icy and refractory material and exhibit dusty comae and volatile outgassing, while asteroids generally lack {vigorously subliming} volatiles and appear as point sources. There are several observational characteristics that are used to separate objects into these populations: detection of a coma, orbit/orbital elements, and nongravitational accelerations. Canonically, comets exhibit comae and nongravitational accelerations due to volatile outgassing \citep{Whipple1950,Whipple1951,Yeomans2004Comets2}. On the other hand, asteroids are characterized as point sources {of light (no extended coma)}, with either purely gravitational accelerations or nongravitational accelerations resulting from radiative processes, such as radiation pressure or the Yarkovsky effect. However, recent advances have shown that this dichotomy oversimplifies the behavior of Solar System small bodies, with the discovery of objects that exhibit observational features that {straddle these} classical classifications. 

So far, two {additional} broad classes of objects {have been identified} along this continuum --- inactive/dormant/extinct comets and active asteroids. Inactive comets have comet-like orbits but exhibit little to no volatile activity. There are several detected subcategories, {such as} the {tailless} ``Manx-comets” on {isotropic} orbits \citep{Jewitt2022}, the Damocloids on Halley-type orbits \citep{Asher1994, Jewitt2005}, and the {so-called} Asteroids on Cometary Orbits (ACOs) on Jupiter Family Comet (JFC) orbits \citep{Jewitt2022}. Damocloids and ACOs are believed to be {highly evolved} comet nuclei that became inactive as a result of cometary fading,  volatile depletion, or mantling \citep{Podolak1985, Prialnik1988,Wang2014,Brasser2015}. {While there is some evidence that the Manx-comets are asteroidal objects that were ejected to the Oort Cloud \citep{Meech2016},} their {dynamical} origins are still unclear, since Jupiter scatters volatile-poor asteroids preferentially into the interstellar medium {rather than} the Oort cloud \citep{Hahn1999, Shannon2015}. Observational evidence indicates that Damocloids and ACOs are likely extinct or dormant comet nuclei \citep{Jewitt2005,Licandro2018}. 

Active asteroids are the inverse --- objects on asteroidal orbits with observable activity \citep{Jewitt2012,Hsieh2017}. For a recent review, see \citet{Jewitt2022}. Main Belt Comets (MBCs) are a class of objects that reside in the main asteroid belt between Mars and Jupiter yet exhibit cometary activity along with a coma \citep{Hsieh2006}. A handful of objects have been identified as MBCs: Comet 133P/(7968) Elst-Pizarro \citep{Elst1996, Boehnhardt1996, Toth2000, Hsieh2004}, 238P/Read, 259P/Garradd, 288P/(30016) 2006 VW139, 313P/Gibbs, 324P/La Sagra, 358P/PANSTARRS, 107P/(4015) Wilson-Harrington, and 433P/(248370) 2005 QN173. Targeted searches for MBCs {suggest} occurrence rates of $<1/500$ and $\sim 1/300$ \citep{Sonnett2011, Bertini2011, Snodgrass2017, Ferellec2022}. See \citet{Jewitt2022} for discussion of the mechanisms responsible for activity in these objects. 

The near-Earth object (3200) Phaethon is of significant interest, exhibiting nongravitational acceleration \citep{Hanus2018} and an association with the Geminid meteoroid stream \citep{Gustafson1989, Williams1993}. Observations have revealed a micron-sized dust production of $\sim 3$ kg s$^{-1}$, too small to explain the association with the Geminids \citep{Jewitt2010, Jewitt2013, Li2013, Hui2017}. Several authors have suggested explanatory mechanisms, including repeated thermally induced stresses \citep{Jewitt2010}, sublimation of minerologically bound sodium \citep{Masiero2021} or iron \citep{Lisse2022}, rotational effects \citep{Ansdell2014, Nakano2020}, and geometric effects \citep{Hanus2016, Taylor2019}. Alternatively, Phaethon's activity may be attributable to sodium sublimation and fluorescence rather than dust \citep{Zhang2023}. {Ultimately, the upcoming DESTINY+ mission will likely clarify the origin of Phaethon's behavior.}

The asteroid (101955) Bennu also exhibits activity, with a measured dust loss rate of $10^{-4}$ g s$^{-1}$ observed by the OSIRIS-REx spacecraft \citep{Lauretta2019,Hergenrother2019,Hergenrother2020}, although it was classified as inactive based on terrestrial observations. However, the source of this activity remains unclear \citep{Bottke2020, Molaro2020, Chesley2020}. There are several nonthermal mechanisms for explaining this sort of activity, {such as micrometeorite} impacts \citep{Snodgrass2010,Bottke2020} and rotational destabilization \citep{Jewitt2014}. 

A potentially distinct population of active asteroids are the ``dark comets” first identified by \citet{Chesley2016} and discussed by \citet{Farnocchia2023} and \citet{Seligman2023}. These are near-Earth objects (NEOs) that exhibit nongravitational accelerations inconsistent with radiative processes, implying the presence of cometary activity. Nevertheless, these objects also have no visible comae, in contrast to the more ``standard” active asteroids. \citet{Farnocchia2023} and \cite{Seligman2023} showed that outgassing is capable of causing the reported accelerations while escaping photometric detection. In addition to the absence of comae, the dark comets exhibit other distinct differences from the active asteroid population --- i) their size, ii) their rotation rate, and iii) their acceleration components. First, these objects are preferentially small, {only} tens of m in radius (with the exception of (523599) 2003 RM which is {a few} hundred m).\footnote{ {The small size, somewhat counterintuitively, exacerbates the issue of  surface dust depletion. This is because molecular and electrostatic forces are more dominant compared to smaller rotational forces on smaller objects \citep{Scheeres2010,Sanchez2020}}.} Their small size may be responsible for the second property, rapid rotation rates. Although less than half of the identified dark comets have measured rotation periods, those periods are remarkably short \citep[{0.046--1.99 h,}][]{Seligman2023}. This may be the result of spin-up due to the YORP effect or outgassing torques, which would operate more effectively  on smaller objects. These objects must be weakly outgassing, since a strongly-outgassing object of these sizes would {likely result in} a destructive rotational disruption cascade \citep{Steckloff2016}. Finally, most of these objects exhibit statistically significant out-of-plane accelerations, which is relatively unexpected. Radiation pressure is generally radial, with minor out-of-plane projections \citep{Vokrouhlicky2000} that do not explain the given accelerations \citep{Seligman2023}. On the other hand, transverse acceleration values are consistent with the Yarkovsky effect or zero acceleration, with the exception of 2003 RM and 2006 RH$_{120}$, which exhibit levels of transverse acceleration incompatible with radiation-induced \citep{Chesley2016,Farnocchia2023}.

One possible explanation for the large out-of-plane acceleration of these objects is preferentially polar outgassing. In this paper, we investigate a seasonal mechanism that can produce this behavior. This mechanism relies on rapid rotation rates to remove equatorial accelerations via symmetry (i.e., the surface rotates sufficiently quickly that we may assume each latitude is at a uniform temperature), and differential heating across the hemispheres to create a net out-of-plane force. This mechanism is broadly similar to the Rotating Jet Model introduced in \citet{Chesley2005}. {However, in this model the combination of surface-covering outgassing and rapid rotation causes the nonpolar force components to average out}. Over the course of an orbit, the subsolar latitude moves across the body (assuming nonzero spin pole obliquity), potentially leading to a net {seasonal} force that can perturb the object's orbit. In this paper, we derive  formulae for the accelerations produced by this seasonal outgassing mechanism, and investigate if this mechanism can explain the accelerations of dark comets. 

{This paper is organized as follows: in Section~\ref{sec:theory}, we derive formulae for the nongravitational acceleration under assumptions {that} produce the balanced outgassing model. In Section~\ref{sec:application}, we apply this model to the dark comets. In Section~\ref{sec:discussion}, we discuss our results and future prospects. Finally, in Appendix~\ref{sec:appendmath}, we derive analytical formulae for the Marsden acceleration components under the balanced outgassing model with a general power-law outgassing scaling. }

\section{Outgassing-Induced Acceleration}\label{sec:theory}

In this section, we derive formulae describing this nongravitational acceleration for an assumed rapidly-rotating and spherical nucleus. In Appendix \ref{sec:appendmath}, we continue this derivation and find formulae for the components of the nongravitational acceleration in these models. While the empirical formulation of \citet{MarsdenV} is also potentially useful, the integrals involved in that problem must be numerically evaluated and are therefore not presented. This derivation is analogous to that presented in  \citet{Vokrouhlicky1998}, with a focus on the sublimative analogue to the seasonal Yarkovsky effect. 

\subsection{Rotational Period Scaling Criteria}\label{subsec:scaling}

In this subsection, we present a scaling criterion that must be met for the outgassing mechanism described in Sec.~\ref{subsec:outgasforce} to apply. Critically, we assume that the rotation of the body is sufficiently rapid that the insolation and temperature, and therefore the outgassing, are essentially constant for a given latitude. For this to be true, the rotation period must be significantly smaller than the thermal relaxation timescale. Note that we also assume that these objects are in the weak sublimation limit, such that cooling from outgassing sublimation is insignificant in comparison to radiative energy loss. This assumption is reasonable as a result of the extremely weak outgassing of the objects under consideration, but is worth noting. {Our equations also assume negligible conduction into the deeper subsurface.} 

Consider a point on the body that has a dayside temperature of $\T_0$. Over the course of the asteroid's rotation, it will reradiate energy to space, with a decrease in the temperature $\Delta \T$. We therefore require $\Delta\T\ll\T_0$, where $\Delta\T$ is lost over a single rotation period.

For a given temperature $\T_0$, the energy radiated in a unit of time is 
\begin{equation}\label{eq:radlost}
    \frac{{\rm d} E}{{\rm d} t}=-\sigma{\epsilon}\T_0^4{\rm d}A\,.
\end{equation}
In Eq.~\eqref{eq:radlost}, ${\rm d}A$ is an area element of the body, {$\epsilon$ is the emissivity of the surface,} and $\sigma$ is the Stefan-Boltzmann constant. The internal thermal energy of the body is given by 
\begin{equation}\label{eq:internalE}
    E{(\T)}=l\,c_p\rho\T{\rm d}A\,.
\end{equation}
In Eq.~\eqref{eq:internalE}, $l$ is the scale depth of the thermal profile, $\rho$ is the density, $c_p$ is the heat capacity per unit mass, and $T$ is the instantaneous temperature of the surface material. Differentiating both sides of Eq. \eqref{eq:internalE} with respect to time gives
\begin{equation}\label{eq:internalEdt}
    \frac{{\rm d} E{(\T)}}{{\rm d} t} = l\,c_p\rho\frac{{\rm d} \T}{{\rm d} t}{\rm d}A\
\end{equation}
We combine Eqs. \eqref{eq:radlost} and \eqref{eq:internalEdt} at $T = T_0$ to find
\begin{equation}\label{eq:Tlost}
    \frac{{\rm d}\T_0}{{\rm d} t} = -\frac{\sigma{\epsilon}\T_0^4}{l\,c_p\,\rho}\,.
\end{equation}
Over a timescale of one rotational period, the temperature lost is found by integrating Eq.~\eqref{eq:Tlost} with respect to time. We also use the fact that, over a timescale $P$, the thermal penetration distance $l\sim\sqrt{\alpha P{/\pi}}$ \citep{Stern1988,Jewitt2017}, where $\alpha$ is the thermal diffusivity and $P$ is the period. Then, we separate variables and integrate over a change of temperature $\Delta T$ and a time period $P$ to find
\begin{equation}
    \Big(1-\frac{\Delta\T}{\T_0}\Big)^{-3}=1+\frac{3\sigma{\epsilon}\T_0^3\sqrt{{\pi} P}}{c_p\rho\sqrt{\alpha}}\,.
\end{equation}
We assume that $\Delta\T\ll\T_0$, and then expand the left-hand side in a Maclaurin expansion,\footnote{If $\Delta\T\ll\T_0$ then both expressions (Equations \eqref{eq:Pmidpoint} and \eqref{eq:Pscaling}) must hold. If $\Delta\T\cancel{\ll}\T_0$, then Eq.~\eqref{eq:Pmidpoint} is incorrect, and the condition given by Eq.~\eqref{eq:Pscaling} is not necessarily true. Therefore, Eq.~\eqref{eq:Pscaling} is necessary but not sufficient for this model's applicability.} finding 
\begin{equation}\label{eq:MacLaurin}
    1+3\frac{\Delta\T}{\T_0}+\mathcal{O}\bigg(\Big(\frac{\Delta\T}{\T_0}\Big)^2\bigg)=1+\frac{3\sigma{\epsilon}\T_0^3\sqrt{{\pi}P}}{c_p\rho\sqrt{\alpha}}\,.
\end{equation}
With some rearranging, this becomes
\begin{equation}\label{eq:Pmidpoint}
    \Delta\T=\frac{\sigma{\epsilon}\T_0^4\sqrt{{\pi}P}}{c_p\rho\sqrt{\alpha}}\,.
\end{equation}
Requiring $\Delta\T\ll\T_0$, {we set the right-hand side of Eq.~\eqref{eq:Pmidpoint} to be $\ll\T_0$.} Using $\alpha=k/(c_p\rho)$, where $k$ is the thermal conductivity, {and assuming that $\epsilon$ has a value close to unity,} we can {solve for the period to find} that we require
\begin{equation}\label{eq:Pscaling}
    P\ll{0.623}\text{ h}\Big(\frac{c_p}{2000 \text{ J kg$^{-1}$ K$^{-1}$}}\Big) \Big(\frac{\rho}{0.5\text{ g cm$^{-3}$}}\Big)\Big(\frac{k}{0.01 \text{ W m$^{-1}$ K$^{-1}$}}\Big) \Big(\frac{280\text{ K}}{\T_0}\Big)^6\,.
\end{equation}
Note that Eq.~\eqref{eq:Pscaling} is identical to Eq.~(8) in \citet{Vokrouhlicky1998} if $\Theta\gg1$, the rapid-rotation limit. For typical minor bodies, $k=0.01$ W m$^{-1}$ K$^{-1}$, $c_p=2000$ J kg$^{-1}$ K$^{-1}$, and $\rho=500$ kg m$^{-3}$ \citep{Gundlach2012,Jewitt2017,Groussin2019}. This corresponds to a thermal diffusivity of $\alpha = 10^{-8}$ m$^2$ s$^{-1}$, which is consistent with literature values \citep{Steckloff2021}.

At 1 au, the effective maximum temperature is $\T_0=280$ K, and so the period $P\ll 1$ h. This is a relatively stringent condition, since the average rotational period of an asteroid is $\sim 6$ hr for asteroids with diameters $D>150$ m \citep{Pravec2002} {and ~10 hr for comets with diameters $D>500$ \citep{Kokotanekova2017}}. This mechanism is therefore likely to only apply to small asteroids, for which a few percent have spin periods $<2$ hr \citep{Pravec2002}. However, this condition is generally satisfied for dark comets with known rotation periods, {which range from 0.046--1.99 h \citep{Seligman2023}}.

\begin{figure}
    \centering
    \includegraphics[width=0.7\linewidth]{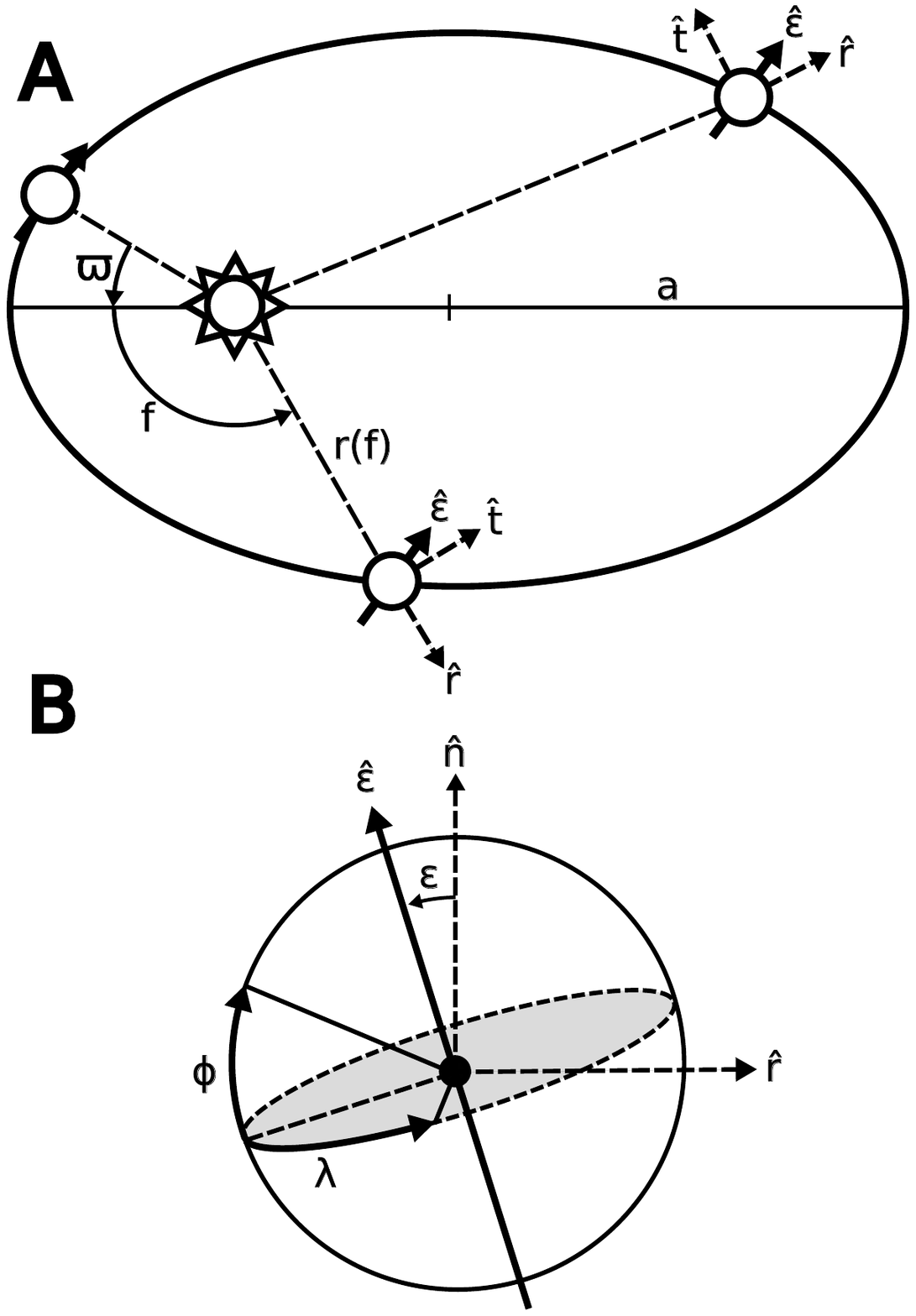}
    \caption{Schematic diagram showing the angles and vectors considered here. A) The orbit of an asteroid and its position at multiple points. The point at the top left is the vernal equinox, with $\varpi$ the angle between the equinox and the perihelion. The vernal equinox is the point at which the object's spin axis is pointed at right angles to the Sun-pointing vector. The true anomaly $f$ is defined to be zero at  perihelion. The heliocentric distance to the object, $r(f)$ is labeled for one of the positions. The solid arrow, which remains constant in the orbit frame, represents the rotation axis, and is labeled as $\boldsymbol{\hat{\varepsilon}}$ in the top right position. Two of the basis vectors, the radial vector $\boldsymbol{\hat{e}}_r$ and the transverse vector $\boldsymbol{\hat{e}}_t$, are shown as dashed arrows. The semimajor axis $a$ is also labeled. B) The asteroid itself, viewed from the orbital plane parallel to the instantaneous velocity. The rotation axis is again labeled $\boldsymbol{\hat{\varepsilon}}$ and the obliquity $\varepsilon$ is shown as the angle between the rotation axis and the out-of-plane vector $\boldsymbol{\hat{e}}_n$. The radial vector $\boldsymbol{\hat{r}}$ is also shown. The gray-shaded region, bound by a dashed line, is the asteroid's equatorial plane. The latitude $\phi$ and the longitude $\lambda$ are shown, measured from the equator and an arbitrary reference line, respectively. }
    \label{fig:angle_diagram}
\end{figure}

\subsection{Outgassing Accelerations}\label{subsec:outgasforce}

In this section, we derive the magnitude of an object's outgassing acceleration at a single instantaneous point on the orbit, albeit with several assumptions. We assume that the object under consideration is i) approximately spherical, ii) outgasses normal to its surface, and iii) {has negligible diurnal outgassing variation. For the purposes of this paper, we will achieve assumption (iii) via rapid rotation }({as defined by }Sec.~\ref{subsec:scaling}). 

We assume that the outgassing acceleration at a single point on an object's surface at an instantaneous orbital position is given by $\boldsymbol{F}(\phi,\lambda,r,f)$. This depends on the latitude $\phi$, the longitude $\lambda$, the heliocentric distance $r$, and the true anomaly $f$. A diagram showing the definition of these angles is provided in Fig.~\ref{fig:angle_diagram}. We assume that the magnitude of the outgassing acceleration can be written as 
\begin{equation}\label{eq:localacc}
    \|\boldsymbol{F}(\phi,\lambda,r,f)\|=C_0\,\frac{L_\sun}{4\pi}Q(\phi,\lambda,f)\,g(r)\,.
\end{equation}
In Eq.~\eqref{eq:localacc}, $C_0$ is a  scaling constant for the acceleration, $L_\sun$ is the Solar luminosity, and $g(r)$ is a Marsden-like function that depends only on the heliocentric distance. This implicitly, and reasonably, assumes that the outgassing rate is related to the insolation magnitude. If $L(\phi, \lambda,r, f)$ is the energy input onto the point $\phi,\lambda$ on the sphere, then we can define $L(\phi,\lambda,r,f)\equiv Q(\phi,\lambda,f)\,L_\sun/(4\pi r^2)$. In Eq.~\eqref{eq:localacc}, we allow $g(r)$ to absorb the $r^{-2}$ from the definition of $L(\phi,\lambda,r,f)$. Therefore, $g(r)$ is a function which contains the entire radial dependence. This equation also serves as a definition of $Q(\phi,\lambda,f)$, which is a factor that describes the fraction of available solar energy input absorbed at a point, and which is given by 
\begin{equation}\label{eq:Qdef}
    Q(\phi,\lambda,f)=(\sin\phi\sin\delta+\cos\phi\cos\delta\cos\lambda)\,.
\end{equation}
In Eq. \eqref{eq:Qdef}, the declination is defined as $\delta=\arcsin(\sin\varepsilon\sin\theta)$, where $\theta$ is an orbital angle defined to be 0 at the vernal equinox. 

As a result of the body's rapid rotation, we assume that the insolation is ``smeared" over lines of constant latitude. We therefore integrate Eq.~\eqref{eq:localacc} over the longitude $\lambda$ to find 
\begin{equation}\label{eq:insolation}
    F(\phi,r,f)=2C_0\,g(r)\,\big[\lambda_0\sin\phi\sin\delta+\cos\phi\cos\delta\sin \lambda_0\big]\,.
\end{equation}
In Eq.~\eqref{eq:insolation}, we have allowed the constant $C_0$ to absorb the constants $L_\sun/4\pi$. The rising longitude $\lambda_0$ is the longitude at which the Sun rises on the asteroid surface for a given latitude, without loss of generality. Since we integrate across all longitudes, the definition of zero longitude is irrelevant. However, for the sake of definiteness, we assume that the Sun rises at $\lambda_0$ and sets at $-\lambda_0$. If perihelion occurs when $\theta=\varpi$, then $\theta=f+\varpi$, with $\varpi$ the vernal equinox angle. The longitude at which the Sun rises, $\lambda_0$, is given by 
\begin{equation}\label{eq:h0}
    \lambda_0=\begin{cases}
    \pi, &\text{for $\tan\phi\tan\delta>1$}\\
    0, &\text{for $\tan\phi\tan\delta<-1$}\\
    \cos^{-1}(-\tan\phi\tan\delta), &\text{otherwise}
    \end{cases}\,.
\end{equation}
In Eq.~\eqref{eq:h0}, the first condition is for latitudes and declinations where the Sun never sets (i.e., summer at high latitudes), the second condition is for circumstances where the Sun never rises (i.e., winter at high latitudes), and the final condition is for intermediate circumstances. Equations \eqref{eq:insolation} and \eqref{eq:h0} are taken from the Appendix of \citet{Berger1978}.

Here, we discuss the reference frame under consideration. We specifically define a reference plane and a reference direction. In this manuscript, we identify the reference plane to be the initial orbital plane of the object under consideration, and set the reference direction to be the vernal equinox of the object. As a result, we assume that the inclination $i=0$. This frame is noninertial in the presence of dynamical evolution of the orbit plane or the spin axis. While we demonstrate in Sec.~\ref{subsec:orbparamvar} that this dynamical evolution is slow over secular timescales, this fact is nevertheless worth noting. It is possible to solve this problem in an inertial reference frame, albeit accompanied by a loss in intuitive understanding of the parameters and an increase in complexity of the result. In the case that dynamical evolution becomes significant in light of future results, a transformation is given in Sec.~\ref{subsec:paramconv}. 

As a result of the longitude-independence, the axial symmetry of the problem ensures that the accelerations perpendicular to the rotation axis cancel and there is only acceleration along the rotation axis. This is a result of the assumption that the rotation is sufficiently rapid such that the nightside temperature is approximately equal to the dayside temperature. With surface-normal outgassing \citep{Steckloff2018,Steckloff2021}, this acceleration is given by $F(\phi,r,f)\sin\phi$. To find the total acceleration along the rotation axis, we integrate this equation over the latitude $\phi$. Due to the piecewise nature of the function over $\lambda_0$, we integrate in three parts --- from $-\pi/2\rightarrow-\phi_0$, $-\phi_0\rightarrow\phi_0$, and $\phi_0\rightarrow\pi/2$. We define $\phi_0$ to be the cutoff at which the domains change in Eq.~\eqref{eq:h0}. By solving for $\tan\phi_0\tan\delta=1$, we find that $\phi_0=\pi/2-\delta$, defining the differences between the domains. Integrating Eq.~\eqref{eq:insolation} by parts over $\phi$, we find that 
\begin{equation}\label{eq:rotaxacc}
    \boldsymbol{F}(r,f)=C_0\,g(r)\,(\boldsymbol{\hat{\varepsilon}}\cdot\boldsymbol{\hat{e}}_r)\,\boldsymbol{\hat{\varepsilon}}\,.
\end{equation}
In Eq.~\eqref{eq:rotaxacc}, $\boldsymbol{\hat{\varepsilon}}$ is the rotation axis, $\boldsymbol{\hat{e}}_r$ is the unit vector of the heliocentric position, and the scaling constant $C_0$ has absorbed {a constant of $\pi^2/2$} from the integration. Eq.~\eqref{eq:rotaxacc} therefore gives the nongravitational acceleration under this model, which is restricted by symmetry to lie along the object's rotation axis. 

\subsection{Parameter Variation}\label{subsec:orbparamvar}

In this section, we address the possibility of variations in the parameters $a$, $e$, $i$, and $\varepsilon$ due to nongravitational accelerations. We compute the secular, first-order time evolutions of these parameters for a nongravitational acceleration of the form given in Eq.~\eqref{eq:rotaxacc}. We present results for an acceleration of the form $g(r)=(1\text{ au}/r)^2$, although we derive a general form with $g(r)=(r_0/r)^\alpha$ in Appendix \ref{sec:paramevo}. Using the results of Appendix \ref{sec:paramevo} (derived from Gauss' planetary equations), we find that
\begin{subequations}\label{eq:simpleparvar}
\begin{align}
    \Big\langle\frac{d a}{d t}\Big\rangle&=0 \label{eq:simpleadot}\\
    \Big\langle\frac{d e}{d t}\Big\rangle&=-\frac{C_0\sin^2\varepsilon}{2na} \Big(\frac{1\text{ au}}{a}\Big)^2  \frac{\eta}{(1+\eta)^2}\,e\sin(2\varpi)\label{eq:simpleedot}\\
    \Big\langle\frac{d i}{d t}\Big\rangle&=\frac{C_0\sin(2\varepsilon)}{4na} \Big(\frac{1\text{ au}}{a}\Big)^2  \frac{e^2\sin(2\varpi)}{\eta^3 \left(1+\eta\right)^2}\label{eq:simpleIdot}
\end{align}
\end{subequations}
In Eq.~\eqref{eq:simpleparvar}, $\eta=\sqrt{1-e^2}$, $n$ is the mean motion and we have also provided the secular evolution of the inclination $\langle di/dt\rangle$. Although this is not a parameter in our outgassing model, since we only need to consider the orbital plane, large values would force the inclusion of other terms in the parameter evolution. Because we are  considering only a single orbit plane, we are free to choose the inclination without loss of generality. Although undefined if $i=0$, we have chosen the argument of pericenter to be equal to the vernal equinox angle $\varpi$, even though $\varpi$ is not technically fixed for a given  orbit. 

Next, we  estimate the magnitude of the variations. For the objects we are considering, $a\sim1$ au, $e\sim0.5$, and $n\sim2\pi$ yr$^{-1}$. The most difficult parameter to estimate is $C_0$. Assuming that the acceleration magnitude $A_i\sim10^{-10}$ {au d}$^{-2}$ \citep{Seligman2023}, we apply Eq.~\eqref{eq:simpleacc} to determine that $C_0\sim A_i$. Setting $\varpi=\pi/2$ to maximize $\sin(2\varpi)$, we find that $\langle\Dot{e}\rangle\sim\langle\Dot{i}\rangle\sim10^{-8}$ yr$^{-1}$. The variation in the semimajor axis $\langle\Dot{a}\rangle$ is identically 0.   The assumption of fixed orbital parameters over the course of one orbit is therefore valid.

There is an additional question, however, of the variation in the obliquity $\Dot{\varepsilon}$. In the absence of outgassing accelerations, the precession of the rotation axis due to the solar gravitational torque has a period of approximately $10^4$--$10^5$ yr \citep{Lhotka2013}, occasionally even shorter if the asteroid's shape
is irregular and rotation period long \citep[e.g.,][but these conditions are {excluded} in our simplified approach]{durech2022}. However, the presence of torques due to outgassing causes spin-up \citep{Jewitt2003b,Drahus2011,Gicquel2012,Maquet2012,Fernandez2013,Steckloff2016,Wilson2017,Eisner2017,Roth2018,Kokotanekova2018,Biver2019,Combi2020,Jewitt2021,Jewitt22}, and can result in significantly more rapid precessional rates \citep{Gutierrez2016,Lhotka2016}. Calculating the change in the obliquity for a general outgassing force requires a knowledge of the outgassing behavior beyond current understanding. In this theory, however, the outgassing jets pass through the object's center-of-mass which fixes the torque to be 0. 

However, an approximation of $\dot{\varepsilon}$ can be found using the notation in \citet{Rafikov2018}, which uses a dimensionless ``lever-arm" parameter $\zeta$ to capture the effective off-center nature of the torquing acceleration. In this formulation, the rate of change of the obliquity (assuming the torque is set to maximally change the obliquity) is 
\begin{equation}\label{eq:epsdot}
    \dot{\varepsilon}=\frac{5}{2}\frac{\zeta A}{D}\tau\,.
\end{equation}
In Eq.~\eqref{eq:epsdot}, we have assumed a spherical object and set $\tau$ the timescale of the torque (roughly the period), $D$ the size of the body, and $A$ the magnitude of the acceleration. The log-averaged value of $\zeta$ for the objects analyzed in \citet{Rafikov2018} is 0.006. Assuming a similar value for the dark comets, and with $D\sim10$ m, $A\sim 10^{-{10}}$ {au d}$^{-2}$, and $\tau=P\sim 0.5$ h, $\dot{\varepsilon}\sim0.1$ rad yr$^{-1}$. This is much more rapid than the precession periods found by \citet{Lhotka2013}, but still sufficiently slow  to be functionally irrelevant over a single orbit.

\section{Application to Dark Comets}\label{sec:application}

In this section, we investigate the applicability of the theory presented in Sec.~\ref{sec:theory} to the nongravitational accelerations of the ``dark comets" described in \citet{Farnocchia2023} and \citet{Seligman2023}. We use {a standard least-squares fit to optical astrometry of these objects, i.e., their observed angular positions in the sky \citep[{see} ][]{Farnocchia2015}. These data are reported in \citet{Farnocchia2023} and \citet{Seligman2023}. Notably, we do not use the standard formulation of \citet{MarsdenV}, but obtain the nongravitational perturbations with Eq.~\eqref{eq:rotaxacc}.}
Instead of $(A_1, A_2, A_3)$, we estimate the orientation of the spin axis and $C_0$.
We assume $g(r) = (1\text{ au}/r)^2$ unless otherwise stated.

\begin{figure*}
    \centering
    \includegraphics[width=0.97\linewidth]{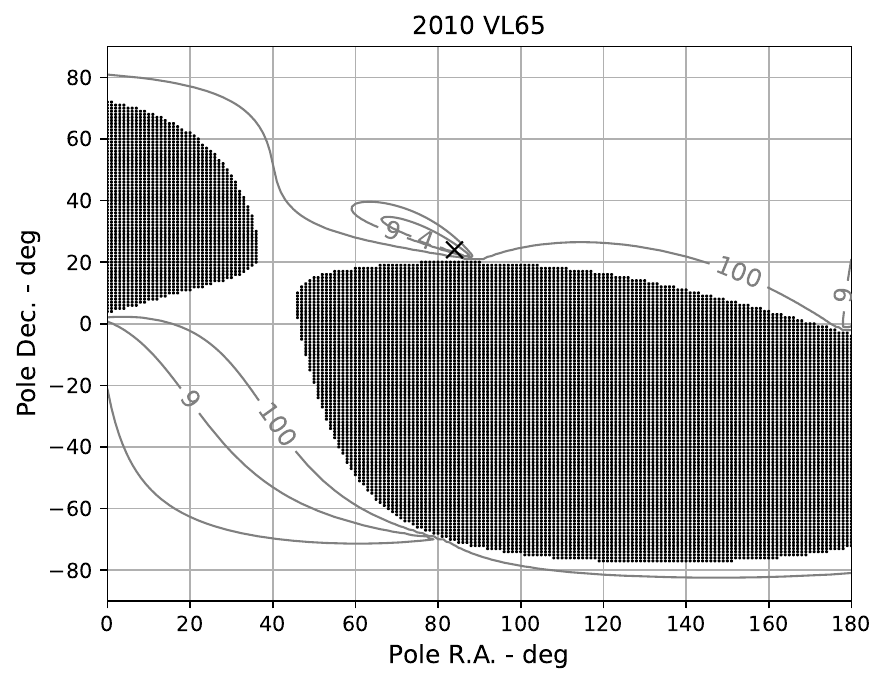}
    \caption{$\chi^2$ of the orbital fit as a function of the pole orientation for 2010~VL65. The cross marks the best fit corresponding to an R.A. of $84^\circ$ and a Dec. of $24^\circ$, resulting in a $\chi^2 = 29.2$ from a fit to 83 observations. The contour levels are $\Delta\chi^2$ levels {relative to the minimum} and the blackened region corresponds to negative values of $C_0$. {The contour in the bottom left indicates where $\Delta\chi^2=9$.}}
    \label{fig:vl5_data_plot}
\end{figure*}

Because of potential nonlinearities and the possible presence of multiple minima, we adopted the same approach of \citet{Chesley2005}. Specifically, we scanned a raster in spin pole's Right Ascension (R.A.) and Declination (Dec.) and for each spin pole we estimated $C_0$ and recorded the $\chi^2$ of the fit.
We discarded pole orientations leading to negative values of $C_0$.
Due to the symmetry of Eq.~\eqref{eq:rotaxacc}, the model cannot distinguish between north and south poles (only returning the orientation of the north-south axis), so we limit the search in R.A. to the interval from $0^\circ$ to $180^\circ$.
Figure~\ref{fig:vl5_data_plot} shows an example of our raster analysis applied to 2010~VL$_{65}$.

For the other objects analyzed by \citet{Seligman2023}, there {is} a single minimum (except for the antipodal symmetry) and Table~\ref{tab:pj_results} shows the best fitting results using the {balanced outgassing} model.
The $\chi^2$ of the fit is always comparable to, if not lower than, the one obtained using the \citet{MarsdenV} model, estimating $(A_1, A_2, A_3)$.
Therefore, the proposed model provides a valid mechanism that allows us to reproduce the astrometric data from nongravitational motion.

\begin{table*}
    \centering
    \caption{\textbf{Best fitting solutions using the {balanced outgassing model}.} We do not include 2006~RH$_{120}$ or 2003~RM, for which the presented model fails to provide an acceptable fit to the data. The $\chi^2$ values for the {balanced outgassing model (BOM)} as well as for the standard \citet{MarsdenV} one are reported. Note that uncertainties are not necessarily linear (see Fig.~\ref{fig:vl5_data_plot}) and so the reported uncertainties are an indication of the uncertainty level, rather than describing a symmetric Gaussian distribution. {This uncertainty is derived by taking the partial derivatives of the residuals with respect to the model parameters, as described in \citet{Farnocchia2015}.} {We do not report reduced $\chi^2$ values, since both models require fitting 3 parameters. Note that 2006~RH$_{120}$ requires  4 parameters to be fit with the addition of the $A_1$ term (see the discussion in the text).}}
    \begin{tabular}{c|ccccc}
         Object & $C_0$ [$10^{-10}$ au/d$^2]$ & R.A. & Dec. & $\chi^2_{{BOM}}$ & $\chi^2_{A1A2A3}$\\\hline
         1998 KY$_{26}$ & $6.5 \pm 1.5$ & $94^\circ \pm 6^\circ$ & $24^\circ \pm 5^\circ$ & $34.4$ & $33.2$\\
         2010 VL$_{65}$ & $3.2 \pm 3.4$ & $84^\circ \pm 7^\circ$ & $24^\circ \pm 5^\circ$ & $29.2$ & $28.2$\\
         2016 NJ$_{33}$ & $1.0 \pm 0.7$ & $36^\circ \pm 12^\circ$ & $15^\circ \pm 3^\circ$ & $36.4$ & $44.6$\\
         2005 VL$_{1}$ & $1.5 \pm 2.3$ & $155^\circ \pm 4^\circ$ & $9^\circ \pm 18^\circ$ & $26.0$ & $26.8$\\
         2010 RF$_{12}$ & $0.9 \pm 0.7$ & $32^\circ \pm 10^\circ$ & $14^\circ \pm 2^\circ$ & $54.1$ & $56.7$\\
         {2006 RH$_{120}$} & {$1.25\pm0.14$} & {$139^\circ\pm3^\circ$} & {$14^\circ\pm4^\circ$} & ${55}$ & ${65.0}$
         \end{tabular}
    \label{tab:pj_results}
\end{table*}

The situation of 2003~RM merits a separate discussion.
Unlike the objects in Table~\ref{tab:pj_results}, 2003~RM is affected by a highly significant transverse acceleration \citep{Farnocchia2023}.
The {balanced outgassing model} fails to reproduce the same net effect of this acceleration and provide an acceptable fit to the data.
We therefore tried to superimpose a transverse acceleration $A_2 (1\text{ au}/r)^2$ and the {balanced outgassing model}.
The improvement over the $A_2$-only fit is $\Delta\chi^2 = 8$, which for the addition of three free parameters to the fit has a $p$-value of 5\%. {This value is the marginal significance, as expressed by the likelihood of obtaining $\Delta\chi^2\geq8$ by adding 3 degrees of freedom.}
This $p$-value is not sufficiently low to justify the addition of new parameters to the model and thus the $A_2$-only model continues to be preferred.

{The situation for 2006~RH$_{120}$ is similarly complex. Specifically, the analytic computations given in Appendix~\ref{sec:appendmath} do not apply in this instance because of the object's geocentric orbit during the observation arc. Eq.~\eqref{eq:rotaxacc} is still applicable, and  it is therefore possible to identify an acceleration pole. However,  the fit is quite poor, with $\chi^2_{{BOM}}=271$ in comparison to $\chi^2_{A1A2A3}=65$. We therefore added an $A_1$ component to the {balanced outgassing} model, consistent with radiation pressure, although the fit with the additional $A_1$ is agnostic to its cause (outgassing or radiation pressure).  Moreover, only the $A_2$ and $A_3$ components are unexpectedly large.  We find a single minimum with $\chi^2=55$ with this updated model. While this provides a better fit than the Marsden model, it is not particularly favored because there is one additional parameter, an $A_1$ acceleration of $(1.6\pm0.09)\,10^{-10}$ au d$^{-2}$. Notably, the jet and the $A_1$ accelerations are of similar magnitude, which may suggest that these accelerations share a similar mechanism. }

\section{Discussion}\label{sec:discussion}

In this paper, we developed a {balanced outgassing} model for seasonal anisotropic outgassing on spherical, rapidly-rotating  bodies that produces a nonzero out-of-plane acceleration. In this model, the rapid rotation of the body cancels nonpolar accelerations via symmetry. Seasonally-varying differential heating over the body's hemispheres results in unbalanced polar accelerations, producing a nonzero acceleration averaged over the course of the eccentric orbit. We demonstrated that this theory is explanatory for most of the nongravitational accelerations of the ``dark comets" identified by \citet{Farnocchia2023} and \citet{Seligman2023}, as this theory produces good fits to the measured nongravitational accelerations for most objects, comparable to the quality of fit using the Marsden parameterization. In general, the rotation poles found for this method are relatively well-constrained, although they are symmetric about the orbital plane.

It is also interesting to note that the {balanced outgassing model} alone cannot explain the acceleration of 2003 RM. The introduction of a transverse acceleration is necessary to fully match observations. This is evidence that 2003 RM is accelerated by the Yarkovsky effect or some similar mechanism, distinct from the remainder of the dark comets, adding to the evidence of its large transverse acceleration and large size. {While 2006~RH$_{120}$ was in a geocentric orbit during the observational arc, it too cannot be explained by balanced outgassing alone. Specifically fitting the observations of this object requires an additional radial component of nongravitational acceleration.}

While this {balanced} seasonal outgassing model is able to reproduce the dark comets' accelerations with high accuracy, we must make several important caveats. Most importantly, this model is highly dependent on {isotropic or negligible diurnal outgassing} as this is invoked to balance nonpolar accelerations; this is therefore ---  admittedly --- somewhat fine-tuned. {Although we invoke rapid rotation in this paper to remove diurnal effects --- justified by rapid rotation periods measured in three dark comets --- other mechanisms could provide a similar diurnal (non)effect; our conclusions therefore do not directly depend on this specific mechanism.} While the dark comets with known rotation periods {likely} satisfy this condition (see Table~1 in \citet{Seligman2023}), the period of the other objects {are presently} unknown. Similarly, this model assumes that the objects under consideration are spherical, which is not necessarily true. {Indeed}, asymmetric objects (e.g., Hartley 2) generally have an aspect ratio close to 2. {Nevertheless}, the only dark comet with a known shape (1998 KY$_{26}$) is roughly spherical \citep{Ostro1999}. Additionally, this model ignores any orbit-scale time delay in the outgassing response to heating.\footnote{Short-term time delay in the outgassing/thermal response is in fact necessary for this mechanism to function, since {this model} requires that the outgassing be smeared out over latitudinal bands.} While this is {likely} not significant {due to the typical dark comet orbital period}, its inclusion will somewhat complicate the mathematics model. 

{It is worth discussing the qualitative effects of failure in the assumptions present in the model, even if quantitative analysis is beyond the scope of this paper. Non-negligible diurnal outgassing would likely produce an increase in the radial acceleration relative to the transverse and out-of-plane accelerations \citep[it is worth noting that diurnal outgassing can also affect the transverse and out-of-plane accelerations, albeit to a much weaker degree based on typical SYORP coefficients ][]{Safrit2021}. This would presumably be due to  a temperature and outgassing differential between the dayside and nightside. The impact of asphericity is highly dependent on the mutual alignment of the rotation relative to the aspherical components.}

There are also several factors to note concerning the application of this model to the dark comets. In this paper, we assume that the acceleration scales as $r^{-2}$, in the absence of evidence concerning the true functional form; {although theoretical work shows that it can scale as $r^{-2.1}$ inside of a species' sublimation radius, which is $\sim$3-4 au for water ice \citep{Steckloff2015}}. Fortunately, {the dark comets orbit within water's sublimation radius, and the differences in the power law index result in acceptable error for the purposes of this work. Furthermore,} the model used here is general and can be used to update these results in response to future measurements. In addition, the parameters derived in this theory are not observed for comparison. \textit{In situ} observations of the rotation characteristics, including the rotation axis, would provide a measurement of these values. 

This model has no bearing on dust production in the outgassing of these objects. However, their small orbital semimajor axes imply near-continual exposure to temperatures sufficient for volatile sublimation over their lifetimes, which may have previously removed surface dust. Regardless of these issues, this model is generally applicable to bodies where outgassing does {not significantly vary over a diurnal cycle}.

The incorporation of additional effects, particularly time-delayed reactions, may help to additionally refine this model. Future data acquisition is also necessary to improve the application of these models, in particular the functional form of the outgassing with respect to the heliocentric distance. The dark comet 1998 KY$_{26}$ is the target of the Hayabusa2 extended mission \citep{Hirabayashi2021,Kikuchi2023}, which will be able to determine the spin pole of 1998 KY$_{26}$ for comparison with our results. \textit{In situ} observations of outgassing will also enable a greater understanding of the mechanism behind these nongravitational accelerations. More sensitive future observations would be necessary to observe the predicted outgassing from these objects, and such observations will provide additional data on the functional form of the nongravitational accelerations. If such information were to become available, the model derived in this paper can be (re)applied to these objects to produce refined parameter estimations. 

{It is  worth noting that the preferential out-of-plane dark comet accelerations  may be the result of selection bias. Out-of-plane accelerations are more readily detectable than in-plane accelerations, which can be mistaken for Yarkovsky or radiation pressure (although the magnitude of the accelerations here is not consistent with those mechanisms). It is possible that the currently-identified dark comets may reflect a larger population of similar objects with more uniform distributions of nongravitational acceleration directions.}

{Assuming that the balanced outgassing model is accurate for dark comets, there are potential implications for their dynamical and formation history. The measured rapid rotations may be an artifact of longer-term outgassing. In addition, the balanced outgassing model requires isotropic outgassing. It is possible that the dark comets are fragments of larger cometary bodies, which is consistent with their small sizes and rapid rotations. }

{This model is quite general, and can be applied to bodies beyond the dark comets. While out-of-plane nongravitational accelerations are rare, the balanced outgassing model may apply to similar accelerations on more traditional active objects. However, such objects are likely not rotating rapidly enough for the outgassing to be fully balanced across the rotation period, since rapid rotation leads to disintegration for larger bodies and the isothermal rotation period is more rapid than the typical disintegration period. Despite this, this model is still relevant to a slow rotator with negligible diurnal outgassing, if such objects exist. In addition, the formulae presented in the Appendix are generally valid for a pair of antipodal outgassing jets, regardless of the object's rotation rate. As a result, this theory may be applicable to a wide variety of active objects.}

Although this seasonal outgassing model is explanatory for the nongravitational acceleration of the dark comets, the paucity of data for these objects implies that this may be a spurious fit. Nevertheless, seasonal {balanced} outgassing {aligned with the spin axis} is a plausible explanation for the out-of-plane nongravitational accelerations on these objects (possibly in addition to other mechanisms, {such as radiation pressure, outgassing jets, etc}). While {the fit quality is similar} to the Marsden formulation, this theory provides a physical explanation for the out-of-plane acceleration exhibited by the dark comets. The general qualitative results of this theory apply to polar outgassing jets and anisotropy, regardless of the mechanism of collimation or asymmetry. Further investigation is necessary in order to refine our understanding of these objects and nongravitational acceleration in general. 

\section*{Acknowledgements}

We thank W. Garrett Levine, Christopher O'Connor, and Steven Chesley for useful conversations and suggestions. {We thank the two anonymous reviewers for their helpful comments.} A.G.T. acknowledges support from the Fannie and John Hertz Foundation {and the University of Michigan's Rackham Merit Fellowship Program}. D.F. conducted this research at the Jet Propulsion Laboratory, California Institute of Technology, under a contract with the National Aeronautics and Space Administration (80NM0018D0004). D.V. acknowledges support from grant 21-11058S of the Czech Science Foundation. {D.Z.S. acknowledges financial support from the National Science Foundation  Grant No. AST-2107796, NASA Grant No. 80NSSC19K0444 and NASA Contract  NNX17AL71A. D.Z.S. is supported by an NSF Astronomy and Astrophysics Postdoctoral Fellowship under award AST-2202135. This research award is partially funded by a generous gift of Charles Simonyi to the NSF Division of Astronomical Sciences.  The award is made in recognition of significant contributions to Rubin Observatory’s Legacy Survey of Space and Time.} J.K.S. acknowledges support from NASA Grant No. 80NSSC19K1313 and NASA Grant No. 80NSSC22K1399.

\appendix

\section{Outgassing-Induced Acceleration Components}\label{sec:appendmath}

\subsection{General Orbit Average}\label{subsec:orbitavg}

In this section, we present a method to average a function over a fixed orbit, which we commonly employ in the remainder of these Appendices. 

We take the orbit average of a function $\Psi(f)$, which is assumed to be an integrable function of the true anomaly $f$. The orbit average is denoted as $\langle\cdot\rangle$. The orbit is characterized by the semimajor axis $a$ and the eccentricity $e$. If the orbit has a period $P$, then 
\begin{equation}\label{eq:orbtimeavg}
    \langle \Psi(f(t))\rangle=\frac{1}{P}\int_0^P\Psi(f(t))\,{\rm d}t\,.
\end{equation}
We therefore must transform Eq.~\eqref{eq:orbtimeavg} to an integral over the true anomaly. Kepler's Second Law gives
\begin{equation}\label{eq:kepler2}
    \frac{df}{dt}=\frac{2\pi}{P} \frac{a^2\eta}{r(f)^2}\, ,
\end{equation}
where $\eta=\sqrt{1-e^2}$. Additionally, the heliocentric distance as a function of true anomaly is given \citep{Murray2000} as 
\begin{equation}\label{eq:radius}
    r(f)=\frac{a \,\eta^2}{1+e\,\cos f}\,.
\end{equation}
We can now rewrite Eq.~\eqref{eq:orbtimeavg} using Equations \eqref{eq:kepler2} and \eqref{eq:radius} to conclude that 
\begin{equation}\label{eq:orbitavg}
\begin{split}
    \langle\Psi\rangle=&\,\frac{1}{2\pi a^2\eta}\int_{-\pi}^{\pi}r(f)^2\,\Psi(f)\,{\rm d}f\\
    =&\,\frac{\eta^3}{2\pi}\int_{-\pi}^{\pi}(1+e\cos f)^{-2}\,\Psi(f)\,{\rm d}f\,.
\end{split}
\end{equation}

\subsection{Parameter Conversions}\label{subsec:paramconv}

In the main body of this text, we report values for an object's spin axis in terms of the right ascension and declination, which are also written as RA, Dec or $\alpha, \delta$ respectively. However, the mathematics for the acceleration components presented in the Appendices are difficult using these values, due to the complexity of relating the geodependent inertial frame (where $\alpha,\delta$ are defined) with the object-dependent orbital frame (where $\varepsilon, \varpi$ are defined). We therefore perform the derivations in the subsequent sections in the orbital frame, but in this section report equations for converting $\alpha,\delta\rightarrow\varepsilon,\varpi$. 

To begin with, we note that $\alpha$ is defined relative to the Earth's vernal equinox, and $\delta$ is defined relative to the terrestrial equator. We will first convert to the ecliptic coordinate system, where the longitude $\lambda$ and the latitude $\beta$ are defined with respect to the ecliptic. Converting from $\alpha,\delta\rightarrow\lambda,\beta$, we find 
\begin{equation}\label{eq:eclipconv}
    \begin{split}
        \sin\beta=&\sin\delta\cos\gamma-\cos\delta\sin\gamma\sin\alpha\\
        \cos\lambda=&\cos\alpha\cos\delta/\cos\beta\\
        \sin\lambda=&(\sin\delta\sin\gamma+\cos\delta\cos\gamma\sin\alpha)/\cos\beta\,.
    \end{split} 
\end{equation}
In Eq. \eqref{eq:eclipconv}, $\gamma$ is the obliquity of the ecliptic with respect to the Earth, and is essentially fixed at $23.4^\circ$. 

Now, we convert from $\lambda,\beta\rightarrow\varepsilon,\varpi$. In the inertial ecliptic coordinate frame $\boldsymbol{\hat{x}},\boldsymbol{\hat{y}},\boldsymbol{\hat{z}}$, the vector direction $\boldsymbol{\hat{s}}$ defined by $\lambda,\beta$ is $\boldsymbol{\hat{s}}=\cos\beta\cos\lambda\,\boldsymbol{\hat{x}}+\cos\beta\sin\lambda\,\boldsymbol{\hat{y}}+\sin\beta\,\boldsymbol{\hat{z}}$. Note that in the inertial frame, $\boldsymbol{\hat{x}}$ points towards the terrestrial vernal equinox, $\boldsymbol{\hat{z}}$ is normal to the ecliptic, and $\boldsymbol{\hat{y}}$ forms a right-handed coordinate system with the two. 

Meanwhile, the orbit-frame coordinate system defined by $\boldsymbol{\hat{x}'},\boldsymbol{\hat{y}'},\boldsymbol{\hat{z}'}$. In this system, $\boldsymbol{\hat{x}'}$ points towards the perihelion point, $\boldsymbol{\hat{z}'}$ is normal to the orbital plane, and $\boldsymbol{\hat{y}'}$ forms a right-handed coordinate system. In this frame, as shown in Fig.~\ref{fig:angle_diagram}, $\varpi$ is the angle between perihelion and the object's vernal equinox, when the spin axis is normal to the radial vector, and $\varepsilon$ is the angle between the spin axis and $\boldsymbol{\hat{z}'}$. Therefore, the spin axis in this frame is written as $\boldsymbol{\hat{s}}=-\sin\varepsilon\sin\varpi\,\boldsymbol{\hat{x}'}+\sin\varepsilon\cos\varpi\,\boldsymbol{\hat{y}'}+\cos\varepsilon\,\boldsymbol{\hat{z}'}$.

Given that these are equivalent vectors, we simply convert between the coordinate systems and then solve the resulting equalities. For convenience, we will write the conversion from the inertial to the orbital system. This conversion is written as $\boldsymbol{\hat{x}'}=x_1\boldsymbol{\hat{x}}+x_2\boldsymbol{\hat{y}}+x_3\boldsymbol{\hat{z}}$, with $y_1,y_2,y_3$ and $z_1,z_2,z_3$ defined similarly. Then 
\begin{equation}\label{eq:inverseeq}
\begin{split}
    x_1=&\phantom{-}\cos\omega\cos\Omega-\cos i\sin\omega\sin\Omega\,, \\
    x_2=&-\cos\Omega\sin\omega-\cos i\cos\omega\sin\Omega\,, \\
    x_3=&\phantom{-}\sin i\sin\Omega\,, \\
    y_1=&\phantom{-}\cos i\cos\Omega\sin\omega+\cos\omega\sin\Omega\,, \\
    y_2=&\phantom{-}\cos i\cos\omega\cos\Omega-\sin\omega\sin\Omega\,, \\
    y_3=&-\cos\Omega\sin i\,, \\
    z_1=&\phantom{-}\sin i\sin\omega\,, \\
    z_2=&\phantom{-}\sin i\cos\omega\,, \\
    z_3=&\phantom{-}\cos i \,.
\end{split}
\end{equation}
In Eq.~\eqref{eq:inverseeq}, $\Omega$ is the longitude of the ascending node, $\omega$ is the argument of perihelion, and $i$ is the inclination. We also introduce three parameters to ease the computation:
\begin{equation}
\begin{split}
    a=&x_1\cos\lambda\cos\beta+x_2\cos\beta\sin\lambda+x_3\sin\beta\,,\\
    b=&y_1\cos\lambda\cos\beta+y_2\cos\beta\sin\lambda+y_3\sin\beta\,,\\
    c=&z_1\cos\lambda\cos\beta+z_2\cos\beta\sin\lambda+z_3\sin\beta\,.
\end{split}
\end{equation}
Therefore, we can set 
\begin{equation}
\begin{split}
    -\sin\varepsilon\sin\varpi=&a\,,\\
    \phantom{-}\sin\varepsilon\cos\varpi=&b\,,\\
    \cos\varepsilon=&c\,.
\end{split}
\end{equation}
We find that
\begin{equation}\label{eq:conveqs}
\begin{split}
    \varepsilon=&\arccos c\,,\\
    \varpi=&\arctan(-a/b)\,.
\end{split}
\end{equation}
This finalizes the conversion from $\alpha,\delta\rightarrow\lambda,\beta\rightarrow\varepsilon,\varpi$.

\subsection{Secular Parameter Evolution}\label{sec:paramevo}

In this section, we derive formulae for the secular first-order variation in the orbital parameters as a consequence of a constant outgassing acceleration of the form given by Eq.~\eqref{eq:rotaxacc}. These can be obtained by orbit-averaging the Gauss planetary equations for semimajor axis $a$, eccentricity $e$, and vernal equinox angle $\varpi$. Although it is initially set at 0, the inclination $i$ must also be considered. The Gauss planetary equations \citep[e.g.,][]{Murray2000,bfv2003} are 
\begin{subequations}\label{eq:parvar}
\begin{align}
    \frac{d a}{d t}&=\frac{2}{n^2a}(\boldsymbol{F}\cdot\boldsymbol{v})\label{eq:adot}\\
    \frac{d e}{d t}&=\frac{\eta^2}{2ae}\frac{d a}{d t}-\frac{\eta}{ane}\Big(\frac{r}{a}\Big)(\boldsymbol{F}\cdot\boldsymbol{\hat{e}}_t) \label{eq:edot}\\
    \frac{d i}{d t}&=\frac{1}{na\eta}\Big(\frac{r}{a}\Big)\cos(f+\varpi)(\boldsymbol{F}\cdot\boldsymbol{\hat{e}}_n) \label{eq:Idot}
\end{align}
\end{subequations}
In Eq.~\eqref{eq:parvar}, $n$ is the mean motion, $\eta=\sqrt{1-e^2}$, and $\boldsymbol{F}$ is the nongravitational acceleration on the object. In Eq.~\eqref{eq:adot}, $\boldsymbol{v}=(na/\eta)(-\sin f\,\boldsymbol{\hat{e}}_P+(e+\cos f)\,\boldsymbol{\hat{e}}_Q)$ is the velocity, where $\boldsymbol{\hat{e}}_P$ is the unit vector pointing to the perihelion, and $\boldsymbol{\hat{e}}_Q$ is a unit vector pointing $90^\circ$ in the forward direction.  Note that while the inclination $i$ is typically 0 without loss of generality --- since we are only considering a single orbit --- if the inclination variation given by Eq.~\eqref{eq:Idot} is large, then adjustments must be made for the nonzero inclination.  

Now there is the question of the acceleration, which is given by Eq.~\eqref{eq:rotaxacc}. We assume that $g(r)=(r_0/r)^\alpha$, where $r_0$ is a constant scaling distance and $\alpha$ is a power-law scaling parameter. In the main body of this manuscript, we use $r_0=1$ au and $\alpha=2$. Here we derive a general form, and then Sec.~\ref{subsec:orbparamvar} quotes the result for that specific circumstance. To fully define Eq.~\eqref{eq:parvar}, we note that the projections of $\boldsymbol{F}$ onto the various basis vectors are given by Eq.~\eqref{eq:axprojs}, with the additional facts that $\boldsymbol{\hat{\varepsilon}}\cdot\boldsymbol{\hat{e}}_P=\sin\varepsilon\sin\varpi$ and $\boldsymbol{\hat{\varepsilon}}\cdot\boldsymbol{\hat{e}}_Q=\sin\varepsilon\cos\varpi$. We use the general averaging formula given by Eq.~\eqref{eq:orbitavg} to find that the time-averaged parameter variations are given by 
\begin{subequations}\label{eq:parvaravg}
\begin{align}
    \Big\langle\frac{d a}{d t}\Big\rangle&=\frac{C_0 r_0^\alpha \sin^2\varepsilon\, e^2\sin(2\varpi)}{8 n a^\alpha}\eta^{2(1-\alpha)}(\alpha+1)(\alpha-2)\,{}_2F_1\Big(\frac{3-\alpha}{2},\frac{4-\alpha}{2};3;e^2\Big)\label{eq:adotavg}\\
    \Big\langle\frac{d e}{d t}\Big\rangle&=-\frac{C_0 r_0^\alpha  \sin^2\varepsilon\, e \sin(2\varpi)}{4 n a^{\alpha+1}}\eta^{2(2-\alpha)}\bigg[\frac{\alpha+1}{2}\,{}_2F_1\Big(\frac{3-\alpha}{2},\frac{4-\alpha}{2};3;e^2\Big)+(3-2\alpha)\,{}_2F_1\Big(\frac{3-\alpha}{2},\frac{4-\alpha}{2};2;e^2\Big)\bigg] \label{eq:edotavg}\\
    \Big\langle\frac{d i}{d t}\Big\rangle&=\frac{C_0 r_0^\alpha  \sin(2\varepsilon)\, e^2 \sin(2\varpi)}{32 n a^{\alpha+1}}\eta^{2(1-\alpha)}(\alpha-3)(\alpha-4)\,{}_2F_1\Big(\frac{5-\alpha}{2},\frac{6-\alpha}{2};3;e^2\Big) \label{eq:Idotavg}
\end{align}
\end{subequations}
By setting $\alpha=2$, one can recover the results presented in Sec.~\ref{subsec:orbparamvar}.

As an aside, ${}_2F_1$ represents the Gauss hypergeometric function, which includes many special functions as limiting cases of its parameter values. This function is defined over $z\in\mathbb{C}$ as 
\begin{equation}\label{eq:hypergeo}
\begin{split}
    {}_2F_1\big(a,b;c;z\big)\equiv&\sum_{n=0}^\infty\frac{\Gamma(a+n)}{\Gamma(a)}\frac{\Gamma(b+n)}{\Gamma(b)}\frac{\Gamma(c)}{\Gamma(c+n)}\frac{z^n}{n!}\\
    =&1+\frac{ab}{c}\frac{z}{1!}+\frac{a(a+1)b(b+1)}{c(c+1)}\frac{z^2}{2!}+\cdots
\end{split}
\end{equation}
In Eq.~\eqref{eq:hypergeo}, $\Gamma$ represents the Euler gamma function. This series is absolutely convergent for $|z|<1$, and is defined on the remainder of the complex plane via analytic continuation.

\begin{figure}
    \centering
    \includegraphics[width=0.6\linewidth]{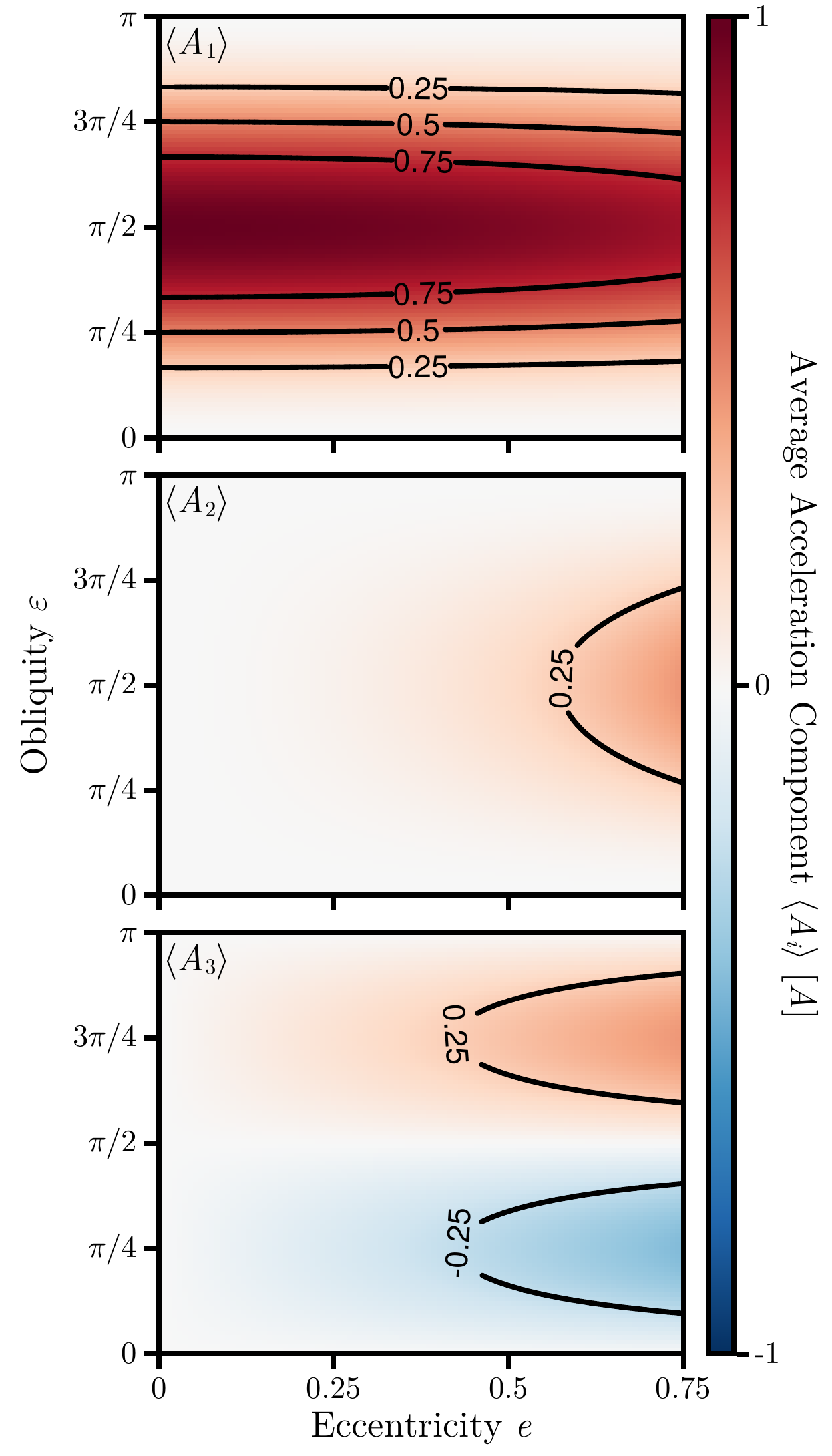}
    \caption{The structure of the acceleration components given by Eq.~\eqref{eq:simpleacc}. We set $\varpi=1$ and show the acceleration in units of the scaling constant $C_0$. While we only show one value of $\varpi$, there is little variation in the structure, with the exception of sign changes --- $A_2$ changes sign at $\varpi=\pi/2,\,\pi,\,3\pi/2$, and $A_3$ changes sign at $\varpi=0,\,\pi$. Note that these values are not shown.}
    \label{fig:accstruct}
\end{figure}

\begin{figure}
    \centering
    \includegraphics[width=0.4\linewidth]{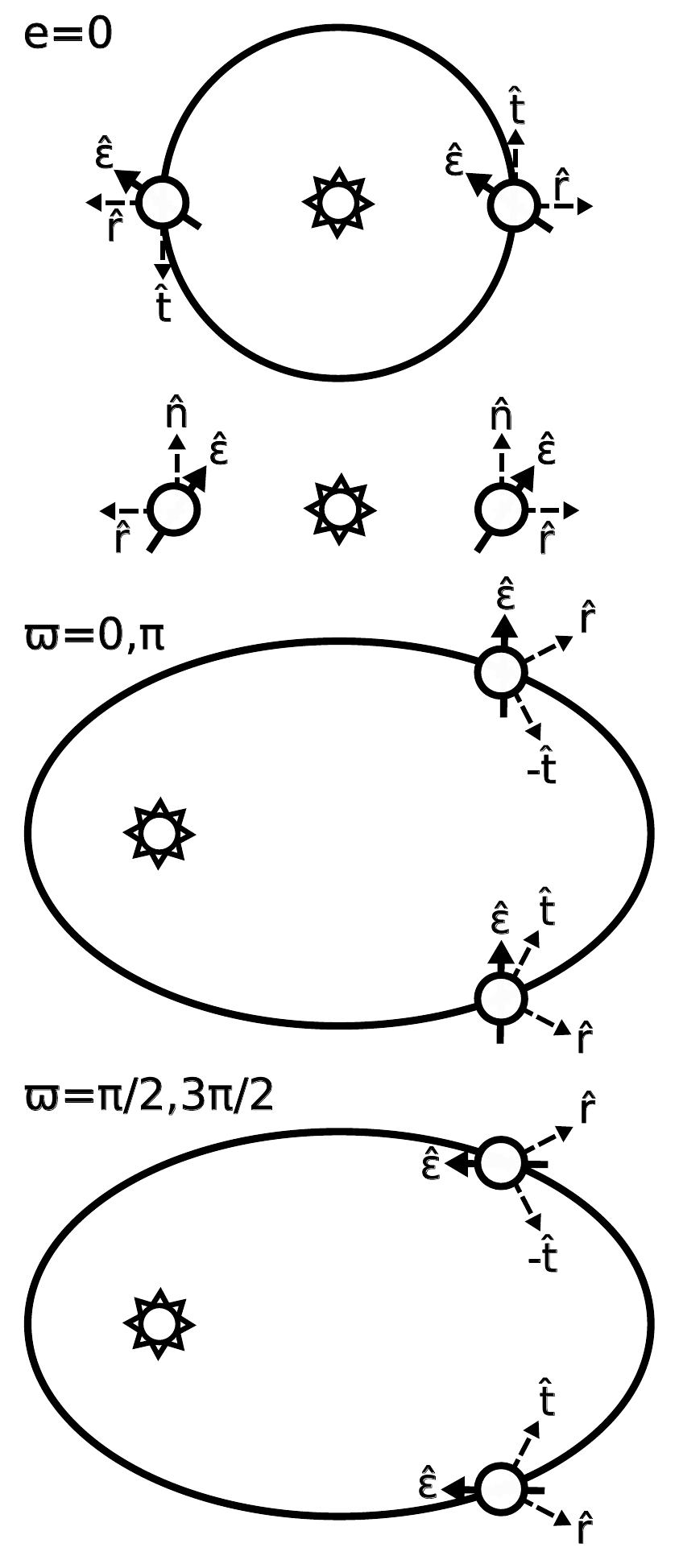}
    \caption{Geometric arguments for the limiting cases given in Table \ref{table:limits}. In all instances, the instantaneous $i$th acceleration component is roughly proportional to the dot product of the Sun-pointing basis rotation axis $\boldsymbol{\hat{\varepsilon}}$ with the $i$th basis vector ($\boldsymbol{\hat{e}}_r,\,\boldsymbol{\hat{e}}_t,\,\boldsymbol{\hat{e}}_n$). The first panel shows the case for $e=0$, a circular orbit. The first part of the panel shows the orbit from above. The $A_1$ component is always in the radial direction, and so $\langle A_1\rangle>0$. While $A_2$ does not have opposite magnitudes at antipodal points, the constant forcing magnitude means that at any point in the orbit, a separate point exists which has the same magnitude but an opposite sign, resulting in cancellation. The second part of the panel shows the orbit from within the orbital plane. This demonstrates that $\langle A_3\rangle=0$. The second panel shows the cases where $\varpi=0,\pi$, which cause cancellation in $\langle A_2\rangle,\,\langle A_3\rangle$  by flipping the direction of $\boldsymbol{\hat{\varepsilon}}\cdot\boldsymbol{\hat{e}}_t,\,\boldsymbol{\hat{\varepsilon}}\cdot\boldsymbol{\hat{e}}_n$. The third panel shows cases where $\varpi=\pi/2,3\pi/2$, which is similar except that $\boldsymbol{\hat{\varepsilon}}$ is rotated by $\pi/2$. }
    \label{fig:geometry_limiting}
\end{figure}

\subsection{Acceleration Component Formulae}\label{subsec:obsaccparams}

The along-axis rotation is difficult to measure for nongravitational accelerations \citep[although it is feasible for some comets under certain models, see][]{Chesley2005}. Instead, we here consider the formulation of \citet{MarsdenV}, which divides the non-gravitational acceleration into three components. In the formulation of \citet{MarsdenV}, we write 
\begin{equation}\label{eq:marsdenform}
\boldsymbol{F}_{\rm NG} = \big( A_1 \boldsymbol{\hat{e}}_r + A_2 \boldsymbol{\hat{e}}_t + A_3 \boldsymbol{\hat{e}}_n\big) \, g'(r)\,.
\end{equation}
In Eq.~\eqref{eq:marsdenform}, the $A_i$'s are constants for a given object which scale the acceleration, $g'(r)$ contains a radial dependence, and the unit vectors point in relevant directions. Specifically, the radial vector $\boldsymbol{\hat{e}}_r$ is radially outward, in the anti-solar direction, $\boldsymbol{\hat{e}}_t$ is transverse to the orbit ``parallel to the line from the Sun to the point in the orbit with true anomaly $90^\circ$ ahead of the comet" \citep{MarsdenII}, and $\boldsymbol{\hat{e}}_n$ is in the out-of-plane direction and forms a right-handed coordinate system with $\boldsymbol{\hat{e}}_r,\,\boldsymbol{\hat{e}}_t$ (see Fig.~\ref{fig:angle_diagram} for clarity). Writing $\boldsymbol{\hat{e}}_i$ to be the $i$th basis vector, the acceleration $\boldsymbol{F}_i=A_i\,\boldsymbol{\hat{e}}_i$. 

In Eq.~\eqref{eq:marsdenform} and in this appendix, we distinguish $g'(r)$ from $g(r)$. $g'(r)$ is the outgassing radial dependence which was assumed when calculating the nongravitational accelerations. On the other hand, $g(r)$ is the outgassing radial dependence in the theoretical model used here. When nongravitational accelerations are derived from orbital fits to astrometric observations for asteroids, $g'(r)=(1\text{ au}/r)^2$ is commonly used \citep[see][]{Farnocchia2023,Seligman2023}. However, comet nongravitational accelerations usually use an empirical formula for $g'(r)$ based on H$_2$O outgassing and given in \citet{MarsdenV}. Since we do not know the true scaling relationship for objects under consideration, we keep $g(r)$, the theoretical scaling, as general as possible. 

In our analysis we  derive the acceleration magnitude $F_i$'s, rather than the $A_i$'s. As a result, caution must be taken to ensure that our analytic results are comparable to observational values. By projecting the acceleration vector onto the $i$th basis, we write Eq.~\eqref{eq:rotaxacc} as 
\begin{equation}\label{eq:acccomp}
    A_i(r, f)=\frac{\boldsymbol{F}\cdot\boldsymbol{\hat{e}}_i}{g'(r)}=C_0\,(\boldsymbol{\hat{\varepsilon}}\cdot\boldsymbol{\hat{e}}_r)\,\big(\boldsymbol{\hat{\varepsilon}}\cdot\boldsymbol{\hat{e}}_i\big)\,\frac{g(r)}{g'(r)}\,.
\end{equation}
In Eq.~\eqref{eq:acccomp}, the $A_i$'s are not constants, but instead depend on the instantaneous orbital position. In order to account for this, we derive the time-average acceleration component over a single orbit. In order to compute this integral using the results of Sec.~\ref{subsec:orbitavg}, we assume that the orbital parameters vary only on secular timescales. In Sec.~\ref{subsec:orbparamvar}, we verify that this assumption holds for the accelerations considered.

In order to find the time-averaged acceleration, we must first find the values of $\boldsymbol{\hat{\varepsilon}}\cdot\boldsymbol{\hat{e}}_i$. The obliquity $\varepsilon$ is defined to be the angle between the rotation axis and the out-of-plane axis $\boldsymbol{\hat{e}}_n$. In addition, the radial and transverse vectors are sinusoidal and periodic over $f$, and the rotation axis $\boldsymbol{\hat{\varepsilon}}$ is offset to be aligned with the transverse vector when $f=-\varpi$ ($\in\mathbb{R}\backslash(2\pi\mathbb{R})$). Note that these angle, and the basis vectors, are all defined in the orbital plane of the body. Therefore 
\begin{equation}\label{eq:axprojs}
    \boldsymbol{\hat{\varepsilon}}\cdot\boldsymbol{\hat{e}}_i=
    \begin{cases}
    \sin\varepsilon\sin(f+\varpi)&i=r\\
    \sin\varepsilon\cos(f+\varpi)&i=t\\
    \cos\varepsilon&i=n\\
    \end{cases}
\end{equation}

Here, we assumed that $g(r)=g'(r)$, so that the true acceleration scaling matches the acceleration scaling assumed in the derivation of the reported $A_i$ values. In Appendix \ref{sec:powerlaw} we derive results for a general power-law scaling, $g(r)= (1\text{ au}/r)^\alpha\,g'(r)$. Combining Equations \eqref{eq:acccomp} and \eqref{eq:orbitavg} and integrating, we find that
\begin{equation}\label{eq:simpleacc}
\langle A_i\rangle=C_0\sin\varepsilon
\begin{cases}
    \sin\varepsilon\big[1-\frac{1+2\eta}{(1+\eta)^2}\,e^2\cos(2\varpi)\big]&i=1\\
    \sin\varepsilon\frac{1+2\eta}{(1+\eta)^2}\,e^2\sin(2\varpi)&i=2\\
    -2\cos\varepsilon\,e\sin\varpi&i=3
\end{cases}
\end{equation}
Observe the similarity to Eq.~(23) in \citet{Vokrouhlicky1998} in the limit of rapid rotation, where $\lambda_1\gg1$. These results are exact and are not constructed via an expansion in eccentricity. 

There are several things to note about this equation. Firstly, $A_1$ is positive, since while $\cos(2\varpi)$ can be negative, the term it leads is always negative and has a strictly smaller magnitude than the leading term. This is physically intuitive, since $A_1<0$ implies that the outgassing is somehow stronger on the hemisphere pointing away from the Sun, which receives lower levels of insolation. In contrast, the signs of $A_2$ and $A_3$ exhibit regular sign flips at specific values of $\varpi$: $A_2$ flips at $\varpi=0,\,\pi/2,\,\pi,\,3\pi/2$ and $A_3$ flips at $\varpi=0,\pi$. The component magnitude increases with eccentricity $e$ and does not depend on the semimajor axis $a$. The lack of $a$-dependence is due to the cancellation of two effects --- the acceleration's scaling with $a^{-2}$ and the time at a given distance's scaling with $a^2$. Note that an $a$-dependence exists if the outgassing does not strictly scale as $r^{-2}$, as we show in Appendix \ref{sec:powerlaw}. Finally, the scaling constant $C_0$ is a leading term in all three components. As a result, the ratios of the acceleration components depend only on the physically relevant parameters $\varepsilon$ and $\varpi$. In Fig.~\ref{fig:accstruct}, we show the structure of the $\langle A_i\rangle$'s in terms of the eccentricity $e$ and the obliquity $\varepsilon$ at a constant $\varpi=1$. 

\subsection{Limiting Cases}\label{subsec:limitcases}

In this section, we evaluate Eq.~\eqref{eq:simpleacc} at several limiting cases for which symmetry allows for exact evaluation, independent of the functional form of the acceleration. This allows us to validate our result by confirming that Eq.~\eqref{eq:simpleacc} reproduces the limiting cases. These cases are presented in Table \ref{table:limits}, and are all confirmed to match the behavior of Eq.~\eqref{eq:simpleacc}. We present diagrams giving geometric arguments for the symmetries in the limiting cases in Fig. \ref{fig:geometry_limiting}.

\begin{table}
    \centering
    \caption{Limiting cases for the acceleration values, which are all confirmed to match the results of Eq.~\eqref{eq:simpleacc}. For each limiting parameter value, the expected sign of the acceleration component is shown. }
    \begin{tabular}{c|ccc}
        Limit & $\langle A_1\rangle$ & $\langle A_2\rangle$ & $\langle A_3\rangle$ \\\hline
        $e\rightarrow0$ & + & 0 & 0 \\
        $\varpi\rightarrow0,\pi$ & + & 0 & 0 \\
        $\varpi\rightarrow\pi/2,3\pi/2$ & + & 0 & +/- \\
        $\varepsilon\rightarrow0,\pi$ & 0 & 0 & 0 \\
        $\varepsilon\rightarrow\pi/2$ & + & +/- & 0 \\
    \end{tabular}
    \label{table:limits}
\end{table}

In all of these cases, it is assumed that the parameters do not meet one of the other presented cases, so each is considered independently. 

\textbf{Case 1:} Consider $e\rightarrow0$, a circular orbit. The Sun-pointing component must be either positive (with a sunward jet) or 0, so $\langle A_1\rangle>0$. The constant heliocentric distance implies that for each point, there exists another with an exactly opposite value of $A_2$. The relationship between antipodal points ensures that $\langle A_3\rangle=0$. 

\textbf{Case 2:} For an elliptical orbit, $\varpi=0,\pi$ implies that the equinoxes occur along the orbit's line of symmetry. Therefore, for every point in the orbit, the acceleration vector will be reversed for the point mirrored across the line of symmetry. Since $\boldsymbol{\hat{e}}_r$ is also mirrored, $\langle A_1\rangle>0$. For $A_2$, the transverse vector will be reflected to its negative, and so the projections will cancel out when averaged. Finally, $\boldsymbol{\hat{e}}_n$ is unchanged, so sign parity means that $\langle A_3\rangle=0$. 

\textbf{Case 3:} If $\varpi=\pi/2,3\pi/2$, then the argument is identical to Case 2, albeit with $\boldsymbol{\hat{\varepsilon}}$ rotated by $\pi/2$ in the orbit plane (see Fig.~\ref{fig:geometry_limiting}). 

\textbf{Case 4:} If $\varepsilon=0,\pi$, then at every point in the trajectory the insolation on both hemispheres will be equal, and the acceleration along the rotation axis will cancel to be $0$. Therefore, there can never be a net acceleration along any axis, and so all components must be 0. 

\textbf{Case 5:} If $\varepsilon=\pi/2$, then the pole is restricted to the orbital plane. Therefore, $\langle A_3\rangle=0$, although $\langle A_1\rangle$ and $\langle A_2\rangle$ can still be nonzero.

\subsection{Power Law Outgassing Scaling}\label{sec:powerlaw}

In this section, we derive the acceleration components for a general power-law force. While this results in a complicated expression, it is analytic and exact, suitable for objects with atypical force scalings \citep[such as 67P,][]{farnocchia2021}. We here assume that $g(r)= (1\text{ au}/r)^\alpha\,g'(r)$. Following the results of Secs. \ref{subsec:orbitavg}, \ref{subsec:obsaccparams}, and \ref{subsec:outgasforce}, we find that we must solve
\begin{equation}
    \langle A_i\rangle=C_0\sin\varepsilon\, \Big(\frac{\text{1 au}}{a}\Big)^{\alpha}\eta^{3-2\alpha}\int_0^{2\pi}\text{d}f(1+e\cos f)^{\alpha-2}\sin(f+\varpi)
    \begin{cases}
        \sin\varepsilon\sin(f+\varpi)&i=1\\
        \sin\varepsilon\cos(f+\varpi)&i=2\\
        \cos\varepsilon&i=3
    \end{cases}
\end{equation} 
These integrate to be 
\begin{equation}\label{eq:powerlawacc}
    \langle A_i\rangle=C_0\,\eta^{3-2\alpha}\Big(\frac{\text{1 au}}{a}\Big)^{\alpha}
    \begin{cases}
        \sin^2 \varepsilon \Big[{}_2F_1\Big(\frac{2-\alpha}{2},\frac{3-\alpha}{2};2;e^2\Big)+\frac{1}{4}e^2\sin^2(\varpi)(\alpha-2)(\alpha-3)\,{}_2F_1\Big(\frac{4-\alpha}{2},\frac{5-\alpha}{2};3;e^2\Big)\Big]&i=1\\
        \frac{1}{8}\sin^2\varepsilon\,e^2 \sin(2\varpi) (\alpha-2)(\alpha-3)\,{}_2F_1\Big(\frac{4-\alpha}{2},\frac{5-\alpha}{2};3;e^2\Big)&i=2\\
        \frac{1}{2}\sin(2\varepsilon)\, e\sin \varpi \,(\alpha-2)\,{}_2F_1\Big(\frac{3-\alpha}{2},\frac{4-\alpha}{2};2;e^2\Big)&i=3
    \end{cases}
\end{equation}

In Eq.~\eqref{eq:powerlawacc}, ${}_2F_1\big(a,b;c;z\big)$ is the Gauss hypergeometric function defined by Eq.~\eqref{eq:hypergeo}. Although complicated, these can be evaluated numerically to high precision. Note that in Sec.~\ref{subsec:obsaccparams}, we use a specific case of Eq.~\eqref{eq:powerlawacc}. As a confirmation our result, Eq.~\eqref{eq:simpleacc} can be obtained from Eq.~\eqref{eq:powerlawacc} by setting $\alpha=0$. Measurements of $\alpha$ values for the dark comets will allow us to obtain robust results for $\varepsilon$ and $\varpi$ using Eq.~\eqref{eq:powerlawacc}.

\bibliographystyle{cas-model2-names}

\bibliography{main}

\end{document}